\documentstyle[aps,amssymb,psfig]{revtex}
\begin{document}

\author{Simonetta Frittelli$^{a,b}$ \and  
	Thomas P. Kling$^{b}$ \and 
	Ezra T. Newman$^{b}$ \\ 
	$^{a}$Department of Physics, Duquesne University, Pittsburgh, 
		PA 15282\\ 
	$^{b}$Department of Physics and Astronomy, University of 
		Pittsburgh,
		Pittsburgh, PA 15260} 
\title{
\rightline{\small{\em To appear in Phys. Rev. D (January 2001)\/} }
	Image distortion from optical scalars in non-perturbative 
	gravitational lensing}
\date{\today} 
\maketitle

\begin{abstract}
 
In a previous article concerning image distortion in non-perturbative
gravitational lensing theory  we described how to introduce shape and
distortion parameters for small sources. We also showed how they could be
expressed in terms of the scalar products of the geodesic deviation
vectors of the source's pencil of rays in the past lightcone of an
observer. In the present work we give  an alternative approach to the
description of the shape and distortion parameters  and their evolution
along the null geodesic from the source to the observer, but now  in
terms of the optical scalars (the convergence and shear of null vector
field of the  observer's lightcone) and the associated optical equations,
which relate the optical  scalars to the curvature of the spacetime. 

\end{abstract}

\section{Introduction}


The distortion of images caused by the gravitational field of a massive
deflector (or ``lens'') is very well understood in the case of weak
fields and thin lenses, where the gravitational field can be treated 
via the linearsuperposition of the fields of point-like masses. In this 
case, imagesappear multiplied, magnified, or distorted, depending on the 
alignmentof the source, the deflector and the observer. We assume the 
reader isfamiliar with the standard, thin-lens theory of gravitational 
lensing asis found in \cite{EFS}.

The standard theory of gravitational lensing provides the necessary
tools for the study of the distortion of images via the linear mapping
$Y=\mbox{\boldmath$A$}X$ where $\mbox{\boldmath$A$}$ is the Jacobian of
the lens map, 

\begin{equation}
	\vec{y}
   =	\vec{F}(\vec{x})
   =	  \vec{x}
	+\frac{D_{ls}}{D_s}\vec{\alpha}(\vec{x})
\end{equation}

\noindent i.e., where $\mbox{\boldmath$A$} = \partial
\vec{F}/\partial\vec{x}$ and where $\vec{x}$ is the angular position of
the image (as seen on the observer's celestial sphere) and $\vec{y}$ is
the angular position of a point source in the background (or unlensed)
spacetime.  $D_l$ and $D_s$ are the distances of the observer from the
lens and source and $D_{ls}$ is the distance between the lens and source.
We denote by $\vec{\alpha}(\vec{x})$ the deflection angle of the lightray
in the lens plane.  Note that both $\vec{x}$ and $\vec{y}$ can be
considered, respectively, as the rescaled position coordinates, $\vec{\xi
}=\vec{x}D_l$ of the image and $\vec{\eta}=\vec{y}D_s$ of the source,  in
the lens and source planes, as represented in Fig.~\ref{fig:distIIa}. 
When no lens is present, the lens equation reduces to
$\vec{y}=\overrightarrow{x}$ and 
$\mbox{\boldmath$A$}=\mbox{\boldmath$I$}$, the identity matrix. If there
is a small, but extended source, we denote a central point of the source
by  $\vec{y}_0$ and any point on its boundary by $\vec{y}$. The central
point is imaged at $\vec{x}_0$, while the image of the boundary point is
at $\vec{x}$. The displacement of the image boundary from its center,
$\Delta \vec{x}=\vec{x}-\vec{x}_0\equiv X$ and that of the source,
$\Delta \vec{y}=\vec{y}-\vec{y}_0\equiv Y$, are related by the Jacobian
of the lens map, i.e., by 

\begin{equation}
	Y = \mbox{\boldmath$A$}X.
\end{equation}

The map is frequently expressed as

\begin{equation}
\mbox{\boldmath$A$}
   =
	\left(\begin{array}{cc}
		1 -\kappa -\gamma_1& -\gamma_2\\
						&\\
		-\gamma_2&\hspace{0.3cm}1 -\kappa +\gamma_1
		\end{array}
	\right)
\label{Astandard}
\end{equation}

\noindent where the quantities $\kappa $ and $\gamma =\sqrt{\gamma
_{1}^{2}+\gamma _{2}^{2}}$ are interpreted, respectively, as the
convergence and shear of the image wih respect to the unlensed source.
[Later we will give a more precise meaning to these quantities in terms
of the convergence and shear of the null geodesic congruence defined by
the observer's past lightcone.] The inverse of $\mbox{\boldmath$A$}$,
i.e., $\mbox{\boldmath$A$}^{-1}$, is referred to as the {\it
magnification matrix\/} and ``carries'' the source's shape into the
image's shape, via  

\begin{equation}
	X=\mbox{\boldmath$A$}^{-1} Y
\end{equation}

We are here interested in obtaining an analog of this approach to
distortion in a generic case that applies regardless of the strength of
the gravitational field and without reference to thin lenses. We refer
to this as a non-perturbative approach to image distortion. 

In this paper, we consider small elliptical sources, which can be
completely described by three {\it shape parameters\/:} area, semiaxes
ratio and semiaxes orientation relative to a fixed direction. They are
dealt with by means of connecting vectors in the null geodesic
congruence that forms the past lightcone of the observer.  The source
and image descriptions in terms of Jacobi fields are developed in full
detail in our preceding paper~\cite{FKNI}, where we obtain expressions
for the three parameters that measure the ``total'' distortion of the
image with respect to the source: the image's solid angle, semiaxes
ratio and orientation as compared to the source's parameters.  

In addition,  we consider the distortion of the pencil of rays of the
elliptical source as it travels towards the observer, where it
eventually defines the image. This is an ``infinitesimal'' distortion;
The distortion of the image is the result of the cumulative effect of
such infinitesimal distortion of the pencil of rays along the null
geodesic from the source to the observer.   

Jacobi fields appear naturally in the context of gravitational
lensing~\cite{penroseoptical,blandfordjacobi,EFS}.
They proved particularly useful in understanding spacetime
singularities (see \cite{HE}, and \cite{penrosespinors}). More
recently, Jacobi fields have been used~\cite{Erice,nsf2,conjugate} to
understand  issues in the approach to the Einstein equations referred to
as null surface formulation~\cite{nsf1,nsf2}. The feature of Jacobi
fields that we focus on in this paper is the fact that the geodesic
deviation equations, governing Jacobi fields, acquire a particularly
simple form in terms of the dynamics (the optical or Sachs equations)
for the optical scalars (convergence, $\rho $, and shear, $\sigma $) of
the null geodesic congruence. 

In this paper we derive a relationship between the distortion of the
pencil of rays of the elliptical source and the optical scalars of the
geodesic congruence.  This relationship applies in arbitrary spacetimes,
without reference to any approximation. Even though our derivation makes
use of a non-perturbative definition of image distortion based on Jacobi
fields, the Sachs equations allow us to eventually eliminate the Jacobi
fields and obtain a relationship purely between the change in shape of
the pencil of rays and the optical scalars.  Via the optical equations,
this can be interpreted as a cause and effect relationship directly
between the curvature of the spacetime and the distortion of the image. 

Here, as in our companion paper~\cite{FKNI}, we assume that the source 
being imaged does not lie across a caustic of the past lightcone of the
observer.

\section{Summary of non-perturbative image distortion}


We summarize, for easy reference, the salient aspects of our approach to
image distortion as developed in~\cite{FKNI}, and as represented in
Fig.~\ref{fig:distIIb} (Readers familiar with~\cite{FKNI} may want to skip 
this section).  A non-perturbative approach to distortion
requires, as a fundamental tool, a non-perturbative lens mapping: a
mapping from the image location to the source location that does not
rely on weak-field nor thin-lens regimes.  As explained
in~\cite{EFNspacetime,FNuniversal,FKNschw}, such a lens mapping can be
obtained, in principle, from the expression

\begin{equation}
	z^a = F^a(z^a_0(\tau), s,\theta,\phi)
\label{lensmap}
\end{equation}

\noindent which gives the coordinates $z^a$ of source points on the past
lightcone of the observer, located on the worldline $z_0^a(\tau)$,  in
terms of the null geodesic that connects the source with the observer.
Eq.~(\ref{lensmap}) can be obtained, for instance,  by integrating the
null geodesic equation, or by solving the eikonal
equation~\cite{eikonalflat,eikonalasymp}. (Lens equations in generic
spacetimes without the thin-lens approximation are also being considered
by other authors~\cite{Perlick1,Perlick2}. In particular, see
\cite{blandfordkerr} for exact lensing in Kerr spacetime.) The angles $(\theta,\phi)$
represent  the direction of the null geodesic at the observer's location
and specify the angular location of the image on the celestial sphere in
standard spherical coordinates, whereas $s$ gives the parameter distance
of the source to the observer along the null geodesic (it can be thought
of as an affine parameter).  The tangent vectors to the null geodesics
in the lightcone are

\begin{equation}
	\ell^a \equiv \frac{\partial F^a}{\partial s}.
\end{equation}

\noindent Associated with each null geodesic, there is a pair of 
parallel propagated spacelike vectors, $(e_1^a,e_2^a)$, which span the
space of spacelike vectors orthogonal to $\ell^a$, and which allow us to
compare angles at two different locations along a null ray.  The
parallel-propagated basis is defined by

 \begin{eqnarray}
	\ell^b\nabla_{\!b} e_1^a &=& \ell^b\nabla_{\!b} e_2^a = 0,\\
	e_1\!\cdot \ell &=& e_2\!\cdot \ell = 0, \\
	e_1\!\cdot e_1 &=& e_2\!\cdot e_2 = -1,	\\
	e_1\!\cdot e_2 &=& 0.			
\end{eqnarray}

\noindent Even though such parallel propagated basis are only defined up
to a fixed rotation, in principle a unique such basis could be picked in
the case that the electromagnetic radiation emitted by the source is
polarized.  In such case, the polarization vector of the radiation is
parallel transported~\cite{EFS} and defines for us one leg, say $e_1^a$,
of our basis.  

The situation that we consider is that of a small source located at the
value $s^*$ along the geodesic.   More specifically, $s*$ represents a
central point in the intersection of the source's worldtube with the
observer's past lightcone. By assumption, the source's worldtube does
not intersect the caustic of the lightcone, therefore the intersection
with the lightcone is continuous and produces a
single (distorted) image. The source's visible shape is defined by this
intersection and, as a set, is connected to the observer by a pencil of
rays. If the source is small, the pencil of rays consists of a bundle of
null geodesics that are neighboring to a fixed null ray from the
source's ``center'' to  the observer. Points on the pencil are thus
reached by connecting vectors (Jacobi fields).  These are solutions to
the geodesic deviation equation. Given two linearly independent
solutions, $(M_1^a(s),M_2^a(s))$, any  other Jacobi field $Z^a$ can be
expressed as

\begin{equation}
	Z^a(s) = \alpha M_1^a(s) + \beta M_2^a(s)
\end{equation}

\noindent where $\alpha$ and $\beta$ do not depend on $s$.  One natural
basis of Jacobi fields is found by taking derivatives of the lens
mapping, in the form

\begin{equation}
	\widetilde{M}_1^a = \frac{\partial z^a}{\partial \theta},
	\hspace{1cm}
	\widetilde{M}_2^a = \frac{1}{\sin\theta} 
			    \frac{\partial z^a}{\partial \phi}.
\end{equation}

\noindent Although all Jacobi fields associated with the observer's past
lightcone  vanish at the observer's location  (the apex of the
lightcone), these two are such that  their $s-$derivatives are 
orthonormal at the observer's location.   As $s$ varies along the null
geodesic,  their components along $(e_1^a, e_2^a)$ form an $s-$dependent
matrix that we refer to as 
$\widetilde{\mbox{\boldmath$J$}}^i_j(s)\equiv \widetilde{M}_j^a e_a^i$. 
We see that, by construction, 
$s^{-1}\widetilde{\mbox{\boldmath$J$}}^i_i(s) \to \delta^i_j$ as ${s\to
0}$.

This matrix evaluated at the source's location $s^*$ defines for us the
{\it Jacobian of the lens mapping\/}, denoted
$^*\!\!\widetilde{\mbox{\boldmath$J$}}$. Namely, the Jacobian of the
lens mapping is 

\begin{equation}
	^*\!\!\widetilde{\mbox{\boldmath$J$}} 
   \equiv  
	\widetilde{\mbox{\boldmath$J$}}(s^*),
\label{consist1}
\end{equation}

\noindent and is its inverse
$^*\!\!\widetilde{\mbox{\boldmath$J$}}^{-1}$ is our {\it magnification
matrix\/}. Notice that, in a weak-field, thin-lens regime, our Jacobian 
\mbox{\boldmath${}^*\!\!\widetilde{J}$} is related to the Jacobian
\mbox{\boldmath$A$} of the thin-lens theory via

\begin{equation}\label{AfromJ}
    \mbox{\boldmath$A$}
   =\frac{\mbox{\boldmath${}^*\!\!\widetilde{J}$}}{s^*},
\end{equation}

\noindent which is due to the fact that we have not scaled our Jacobi
fields at the source's location. (In a generic spacetime, i.e., in the
absence of a flat background, there is no geometric meaning to such a
scaling.) 

For ease of describing the source's shape, it is convenient to use a
basis of Jacobi fields $(M_1^a, M_2^a)$, rather than
$(\widetilde{M}_1^a, \widetilde{M}_2^a)$, that is identical  to the
parallel propagated basis $(e_1^a, e_2^a)$ at the source's location. See
Fig.~\ref{fig:distIIc}. We introduce  the notation that any quantity
defined or evaluated at the source is preceded by a ${}^*$.   We thus
have

\begin{eqnarray}
	{}^*\!M_1^a &=& {}^*\!e_1^a	,\label{m1*}\\
	{}^*\!M_2^a &=& {}^*\!e_2^a	.\label{m2*}
\end{eqnarray}

\noindent In terms of $(\widetilde{M}_1^a,\widetilde{M}_2^a)$ we have
that 

\begin{equation}
      M_i^a 
   = (^*\!\!\widetilde{\mbox{\boldmath$J$}}^{-1})^j_i
	    \widetilde{M}_j^a(s)
\end{equation}

\noindent If we define the components of $M_i^a$ in the parallel
propagated basis $e_i^a$ by

\begin{equation}\label{jay}
	\mbox{\boldmath$J$}^i_j (s)= M_j^a(s)e_a^j,
\end{equation}

\noindent where $e^j_a$ is the dual to $e_j^a$, we see immediately that 

\begin{equation}
	\mbox{\boldmath$J$}^i_j (s^*)
   =   ({}^*\!\!\widetilde{\mbox{\boldmath$J$}}^{-1})^k_j
	{}^*\!\widetilde{M}_j^a{}^*\!e^i_a
   =   ({}^*\!\!\widetilde{\mbox{\boldmath$J$}}^{-1})^k_j
	{}^*\!\!\widetilde{\mbox{\boldmath$J$}}^i_k
   =\delta^i_j,
\end{equation}

\noindent as required by Eqs.~(\ref{m1*})-(\ref{m2*}).  One also sees,
from  $\lim_{s\to 0} s^{-1}\widetilde{\mbox{\boldmath$J$}}^i_j(s) =
\delta^i_j$, that 

\begin{equation}
	\lim_{s\to 0} \frac{1}{s}\mbox{\boldmath$J$}^i_j(s) 
   = ({}^*\!\!\widetilde{\mbox{\boldmath$J$}}^{-1})^i_j,
\label{consist2}
\end{equation}

\noindent an important result that allows us to map the source's shape
to the image's shape on the observer's celestial sphere. This result is
fundamental to the construction in this paper, since it allows us  to
obtain the magnification matrix by continuously moving along the null
geodesic from the  source to the observer. 

It is important to notice that the vectors $(e_1^a, e_2^a)$ can be
expressed in terms of  $(M_1^a, M_2^a)$, by 

\begin{eqnarray}
	e_1^a &= &  \alpha \cos(\lambda+\nu) M_1^a
		   +\beta \cos\lambda M_2^a  ,\label{e1} \\
	e_2^a &= &  \alpha \sin(\lambda+\nu) M_1^a
		   +\beta \sin\lambda M_2^a  ,\label{e2}
\end{eqnarray}

\noindent where

\begin{eqnarray}
	\alpha 
   &=&  \sqrt{\frac{-M_2\!\cdot\!M_2}
	     {  M_1\!\cdot M_1 M_2\!\cdot M_2
	      -(M_1\!\cdot M_2)^2}}		\label{alpha}\\
	\beta 
   &=&  \sqrt{\frac{-M_1\!\cdot\!M_1}
	     {  M_1\!\cdot M_1 M_2\!\cdot M_2
	      -(M_1\!\cdot M_2)^2}}		\label{beta}\\
	\cos\nu 
   &=&  \frac{M_1\!\cdot\!M_2}
	     {\sqrt{M_1\!\cdot\!M_1 M_2\!\cdot\!M_2}} \label{nu}
\end{eqnarray}

\noindent and $\lambda$ is determined, up to a constant, by

\begin{equation}\label{lambdadot}
	\dot{\lambda} 
   = 	\frac{       M_1\!\cdot\!M_2}
	     {2\sqrt{  M_1\!\cdot M_1 M_2\!\cdot M_2
		     -(M_1\!\cdot M_2)^2}}
	\frac{d}{ds}\log\left(
		\frac{M_2\!\cdot\!M_2}{M_1\!\cdot\!M_2}
			\right).
\end{equation}

\noindent At the source, by Eqs.~(\ref{m1*})-(\ref{m2*}), we have

\begin{equation}
	{}^*\!\alpha = 1,	\hspace{1cm}
	{}^*\!\beta = 1,	\hspace{1cm}
	{}^*\!\nu= -\frac{\pi}{2}	,	\hspace{1cm}
	{}^*\!\lambda= \frac{\pi}{2}.
\label{starredvalues}
\end{equation}  

A small elliptical source at $s^*$ of semiaxes $L_+$ and $L_-$, oriented
so  that the semimajor axis lies at an angle $\delta_*$ with
${}^*\!e_1^a$, is described  parametrically by

\begin{equation} 
{}^*Y^a(t) = Y^1(t)\,{}^*\!e_1^a+Y^2(t)\,{}^*\!e_2^a 
           = Y^1(t)\,{}^*\!M_1^a+Y^2(t)\,{}^*\!M_2^a ,
\end{equation}

\noindent where the components $Y^i(t)$ (for $t\in[0,2\pi]$) can be
specified as

\begin{eqnarray}
	Y^1
   &=& \sqrt{\frac{{}^*\!{\cal A}}{{}^*\!{\cal R}}}
	\left({}^*\!{\cal R}\cos t\cos\!{}^*\!\delta 
		   - \sin t\sin\!{}^*\!\delta\right),\\
	Y^2
   &=& \sqrt{\frac{{}^*\!{\cal A}}{{}^*\!{\cal R}}}
	\left({}^*\!{\cal R}\cos t\sin \!{}^*\!\delta 
		    + \sin t\cos\!{}^*\!\delta\right).
\end{eqnarray}

\noindent Here ${\cal R}$ is the ratio of the semiaxes (${\cal R} \equiv
L_+/L_-$)   and ${\cal A}$ is referred to as the area of the ellipse,
even though it is defined simply  as the product of the semiaxes (${\cal
A}\equiv L_+L_-$).  We can follow the lightrays that connect each point
of this source with the observer by defining  the connecting vector
$Z^a(s,t)$ as

\begin{equation}
	Z^a(s,t) \equiv Y^1(t) M_1^a(s) + Y^2(t) M_2^a(s)
\end{equation}

\noindent The image is thus obtained by projecting $Z^a(s,t)$ along the
directions  $(e_1^a,e_2^a)$, in the limit as $s\to 0$:

\begin{equation}
	X^i(t) \equiv \lim_{s\to 0} \frac{1}{s} Y^j(t)M_j^a(s)e_a^i
   =    ({}^*\!\!\widetilde{\mbox{\boldmath$J$}}^{-1})^i_j Y^j(t)
\end{equation}

In order to study the distortion of the image, compared to the shape of
the source, we define the  {\it shape parameters} to be the area of the
ellipse ${\cal A}(s)$, the ratio of its semiaxes ${\cal R}(s)$  and its
orientation $\delta(s)$, for each $s$ as the ellipse is carried by the
null geodesics towards the observer.   Defining, for short, the
following

\begin{eqnarray}
     a(s)
  &=&
	{}^*\!{\cal R}^2\big((\cos\!{}^*\!\delta)^2 M_1\!\cdot M_1
	+(\sin\!{}^*\!\delta)^2 M_2\!\cdot M_2
	+2\sin\!{}^*\!\delta\!\cos{}^*\!\delta M_1\!\cdot M_2 \big) ,	
\label{a}\\
     b(s)
  &=&
       \frac12(1-{}^*\!{\cal R}^2)(M_1\!\cdot M_1+M_2\!\cdot M_2)
      +\frac12(1+{}^*\!{\cal R}^2)\cos(2{}^*\!\delta)
			       (M_2\!\cdot M_2-M_1\!\cdot M_1)\nonumber\\
   &&
	-(1+{}^*\!{\cal R}^2)\sin(2{}^*\!\delta)M_1\!\cdot M_2  , 
\label{b}\\
     c(s)
  &=&
	{}^*\!{\cal R}\Big(
	        \sin(2{}^*\!\delta)(M_2\!\cdot M_2-M_1\!\cdot M_1)
	       +2\cos(2{}^*\!\delta) M_1\!\cdot M_2  
		  \Big)  ,				  
\label{c}
\end{eqnarray}

\noindent it can be seen~\cite{FKNI} that the shape parameters can
be obtained from the Jacobi vectors  $(M_1^a, M_2^a)$ as 

\begin{eqnarray}
       {\cal A}
   &=&
 {}^*\!\!\!{\cal A}  \Big(M_1\!\cdot M_1 M_2\!\cdot M_2
		    -(M_1\!\cdot M_2)^2\Big)^{1/2}	\label{area}\\
{\cal R} &=& \left(
			\frac{2a+b- \sqrt{b^2+c^2}}
			     {2a+b+ \sqrt{b^2+c^2}}
		   \right)^{1/2}			\label{ratio}\\
 \tan\delta
   &=&
	\frac{ \alpha\cos(\lambda+\nu)
		({}^*\!{\cal R}\cos t_+\sin\!{}^*\!\delta 
			+ \sin t_+\cos\!{}^*\!\delta)
	      -\beta\cos\lambda
	  	({}^*\!{\cal R}\cos t_+\cos\!{}^*\!\delta 
			-\sin t_+\sin\!{}^*\!\delta)}
	     { \beta\sin\lambda
		({}^*\!{\cal R}\cos t_+\cos\!{}^*\!\delta 
			-\sin t_+\sin\!{}^*\!\delta)
	      -\alpha\sin(\lambda+\nu)
		({}^*\!{\cal R}\cos t_+\sin\!{}^*\!\delta
			+ \sin t_+\cos\!{}^*\!\delta)}
\label{delta}
\end{eqnarray}

\noindent where $t_+$ is the value of $t$ that corresponds to the
major semiaxis of the ellipse, which is given by  
\begin{equation}
	\cos^2\! t_+ = \frac12\left(1+\frac{1}{\sqrt{1+(c/b)^2}}\right).
\label{extreme}
\end{equation}

\noindent [Other authors~\cite{Panov} have studied the rotation of 
the image of an elliptical source, corresponding to our
Eq.~(\ref{delta}), but in a different physical context 
(the appearance of the polarization of radio emission by sources at
high redshift).]

\noindent It is useful to define, as well, the angular analog of the
area of the ellipse, namely, the  solid angle of the pencil of rays:

\begin{equation}
	\Omega(s) \equiv \frac{\cal A}{s^2}
\end{equation}

The image subtends a solid angle $\Omega_I$, and its semiaxes have a
ratio ${\cal R}_I$  and orientation $\delta_I$ given by

\begin{eqnarray}
       \Omega_I
   &=& \lim_{s\to 0}\frac{{\cal A}(s)}{s^2} = \lim_{s\to 0}\Omega(s),
\label{imarea}\\
	{\cal R}_I 
   &=&  {\cal R}(0),
\label{imratio}\\
 \delta_I
   &=&  \delta(0).
\label{imdelta}
\end{eqnarray}

\noindent The distortion of the image with respect to the source is
measured by the  distortion parameters ${ \frak D}_{\Omega},{\frak
D}_{\cal R}$ and  ${\frak D}_{\delta}$ defined as

\begin{eqnarray}
	{\frak D}_{\Omega} 
&\equiv& \lim_{s\to 0} \left(\frac{s^*}{s}\right)^4
			\frac{{\cal A}^2}{{}^*\!\!{\cal A}^2} 
			- 1
   =	\lim_{s\to 0}
			\frac{\Omega^2}{{}^*\Omega^2} 
			- 1, 				\label{distA}\\
	{\frak D}_{\cal R} 
   &\equiv& \left(\frac{{\cal R}(0)}{{}^*\!{\cal R}}\right)^2 
			- 1,			       \label{distR}\\
{\frak D}_{\delta} &\equiv& \delta(0) - {}^*\!\delta.	\label{distdelta}
\end{eqnarray}	

\noindent If any of the distortion parameters is non vanishing, the image is
said to be distorted with respect to the source. The 
reader is referred to~\cite{FKNI} for complete details of the construction 
in this Section.  

As defined, the distortion parameters are expressed in terms of the Jacobi
basis $(M_1^a, M_2^a)$.  Even though it might be tempting to colloquially
refer to ${\frak D}_{\Omega}$ as the convergence, and to ${\frak D}_{\cal
R}$ as the shear of the image, here we are interested in obtaining  a
relationship between the distortion parameters and the proper convergence
$\rho$ and  shear $\sigma$ of the lightcone. In the remainder of this paper
we show that the dependence on the Jacobi fields can, in fact, be
eliminated in favor of the convergence $\rho$ and shear $\sigma$ of the
lightcone.   

Before we demonstrate the relationship between the distortion parameters
and the optical scalars in full generality, we take a preliminary step. We
develop the relationship in the case of a thin lens treated
non-perturbatively.  This amounts to applying the formalism as developed in
this section to the case that the curvature of the spacetime along one
light ray is a Dirac-delta function.

\section{Thin lenses treated non-perturbatively}


In this Section, we assume that the spacetime is given, and that it is flat
everywhere in the region of interest except at a surface, that will be
interpreted as a lens ``plane''. We will not be concerned with the geometry
of this surface, which in fact may or may not be a plane. We actually
restrict attention to the spacetime in the neighborhood of a given lightray
that connects a small source to the observer, and assume that the situation
is generic to the observer's lightcone, except  in the neighborhood of a
caustic.  This situation can be modeled with a distributional curvature
tensor along the given lightray, with support at the lens plane. 

Jacobi fields $Z^a$ are solutions to the geodesic deviation
equation  

\begin{equation} 
     \ell^c \nabla_c \ell^b \nabla_b Z^a 
   = 
	R_{~bcd}^a \ell^b Z^c \ell^d
\label{jacobieqn} 
\end{equation}
 
\noindent along a fixed null ray with tangent $\ell^a$. This is an
ordinary second-order differential equation for the components of the
Jacobi field, where the curvature of the spacetime along the ray is
considered as {\it given}. Because the space of solutions is
two-dimensional (in the case of {\it null\/} geodesics), then this is
essentially a problem of two equations for two components.  In the
following, we summarize a standard choice~\cite{penrosespinors,Erice}
for the reduction of ({\ref{jacobieqn}) to a linear problem in two
dimensions. 

Firstly, a parallel propagated basis transverse to the null geodesic
congruence is given.  This is the basis $(e_1^a,e_2^a)$ introduced in
the previous Section. However, it is customary to ``complexify'' the
basis, mainly for easier handling and notation.  We define

\begin{equation}
	m^a\equiv \frac{e_1^a+i e_2^a}{\sqrt{2}}
\label{ma}
\end{equation}

\noindent which consequently satisfies $m\!\cdot\!\bar{m}=-1$ and
$m\!\cdot\!\ell=m\!\cdot\!m=\ell^b\nabla_{\!b}m^a=0$.   We consider
$(m^a,\bar{m}^a)$ as our parallel propagated basis.   Similarly, we
arrange our basis of Jacobi fields $(M_1^a,M_2^a)$ into a complex basis

\begin{equation}
	M^a \equiv \frac{M_1^a+i M_2^a}{\sqrt{2}}
\label{MM}
\end{equation}

\noindent which can now be expressed in terms of $(m^a, \bar{m}^a)$ by
means of two complex components $(\xi,\eta)$ via

\begin{equation}
     M^a \equiv \xi m^a + \eta \bar{m}^a 
\label{M1} 
\end{equation}

\noindent (Here, $\xi$ and $\eta$ should {\it not\/} be confused with
the source location on the source plane and the image's location on the
lens plane, which, as explained in Section I, is the notation used
in~\cite{EFS}, unfortunately).  Substituting Eq.~(\ref{M1}) into
Eq.~(\ref{jacobieqn}) and contracting with $\bar{m}^a$ we obtain

\begin{equation}
	D^2 \xi = - (\Phi_{00} \xi + \bar\Psi_0 \eta),
\end{equation}

\noindent where $D\equiv \ell^a\nabla_a$,   $\Phi
_{00}=\frac12R_{ab}\ell^a\ell^b$ and  $\Psi_0=
C_{abcd}\ell^am^b\ell^cm^d$. Similarly, contracting with $m^a$ yields

\begin{equation}
	D^2 \eta = - (\Psi_0 \xi + \Phi_{00} \eta).
\end{equation}

We can arrange these two equations into matrix form as

\begin{equation} 
	D^2 \mbox{\boldmath $X$} = - \mbox{\boldmath $QX$}, 
\label{eqn1} 
\end{equation}

\noindent where {\boldmath $X$} is a matrix containing the components of
the connecting vectors

\begin{equation} 
    \mbox{\boldmath $X$}
   = 
	\left(  \begin{array}{cc} 
		\xi & \bar\eta \\
		\eta & \bar\xi
	 	\end{array} 
	\right),  \label{X} 
\end{equation}

\noindent and {\boldmath $Q$} is a matrix containing the relevant
components of the curvature:

\begin{equation} 
      \mbox{\boldmath $Q$} 
   = 
	\left( \begin{array}{cc} 
	         \Phi_{00} & \bar\Psi_0 \\
		 \Psi_0    & \Phi_{00} 
		\end{array} 
	\right).  
\label{Qdef} 
\end{equation}

\noindent Equation~(\ref{eqn1}), which is equivalent to Eq.~(\ref{jacobieqn}),
represents an alternative  way to obtain Jacobi fields, by integration
from given free initial values, rather than by  derivation of the
lightcone.  However, there is still an alternative version of
(\ref{eqn1})  for Jacobi fields which is a first-order formulation
involving the optical scalars, rather than the curvature.  To obtain
this formulation, we notice that all connecting vectors satisfy

\begin{equation}
	\ell^a\nabla_aZ^b = Z^a\nabla_a\ell^b,
\label{lie}
\end{equation}

\noindent namely, they are Lie-dragged along the congruence. Putting
Eq.~(\ref{M1}) into Eq.~(\ref{lie}) and contracting alternatively with
$m^a$  and $\bar{m}^a$ we obtain

\begin{eqnarray}
	D\xi &=& -(\rho\xi + \bar\sigma\eta)	\\
	D\eta &=& -(\sigma\xi + \rho\eta)
\end{eqnarray}

\noindent with 

\begin{eqnarray}
     \rho &\equiv& m^a\bar{m}^b\nabla_b\ell_a	\label{rhodef}\\
     \sigma &\equiv& m^a\bar{m}^b\nabla_b\ell_a  \label{sigmadef}
\end{eqnarray}

\noindent which is equivalent to the matrix equation

\begin{equation}
	D\mbox{\boldmath$X$} = - \mbox{\boldmath$PX$}
\label{eqn3}
\end{equation}

\noindent with

\begin{equation}
     \mbox{\boldmath$P$} 
   \equiv 
	\left(\begin{array}{cc}
		\rho & \bar\sigma	\\
		\sigma & \rho
		\end{array}
	\right).
\end{equation}

\noindent Eq.~(\ref{eqn3}) is a first-order differential equation for the
Jacobi fields in terms of the optical scalars: the convergence $\rho$,
and the shear $\sigma$, of the geodesic congruence $\ell^a$. As for
{\boldmath$P$} itself, it can be obtained directly from the curvature by
virtue of the Sachs equations for the optical scalars, namely, 

\begin{equation}
	D\mbox{\boldmath$P$} 
      =  \mbox{\boldmath$Q$} + \mbox{\boldmath$P$}^2,
\label{eqn2} 
\end{equation}

\noindent which is obtained by taking a derivative of Eq.~(\ref{eqn3}) and
using Eqs.~(\ref{eqn1}) and (\ref{eqn3}) to eliminate {\boldmath$X$} from
the resulting equation. Thus we think of the optical scalars $\rho$ and
$\sigma$ as {\it given\/}, for the purposes of solving for the Jacobi
fields in Eq.~(\ref{eqn3}). 

We now specialize our discussion to the particular case of a thin lens. 
Namely, the curvature {\boldmath$Q$} is zero outside of a single,
two-dimensional, spatial surface.  Under this assumption, there is only
one value of $s$ along all relevant null rays in the past 
lightcone of the observer where the value of the curvature is non-zero, 
i.e., the point where the null geodesic intersects the lens ``plane''.  
This value will depend on the choice of $(\theta ,\phi)$ labeling the 
null geodesic but we will simply call this value $s=L<0$.  For a given 
ray, the curvature matrix may be written as

\begin{equation} 
     \mbox{\boldmath$Q$}
   = \left( \begin{array}{cc} 
	    {\widetilde{\Phi}}_{00} &
	   |{\widetilde{\Psi}}_{0}|\,e^{-i\,\varphi} \\
	   |{\widetilde{\Psi}}_{0}|\,e^{i\,\varphi} & 
	    {\widetilde{\Phi}}_{00} 
	   \end{array}
     \right) \delta (s - L) 
   \equiv 
	\mbox{\boldmath$T$} \delta (s - L), 
\label{Tdef} 
\end{equation}

\noindent where $({\widetilde{\Phi}}_{00}, |{\widetilde{\Psi}}_{0}|,
\varphi)$ are real constants. 

In the spirit of our current program, the basis $(M_1^a, M _2^a)$ that
we wish to solve for lies on the past lightcone of the observer and
coincides with $(e_1^a,e_2^a)$ at $s^*$. Translating these two conditions
into conditions for \mbox{\boldmath$X$}, by Eqs.~(\ref{M1}), (\ref{MM}) and
(\ref{ma}), we obtain two boundary
conditions on the solutions to Eq.~(\ref{eqn1}), respectively of the form

\begin{eqnarray}
	\mbox{\boldmath$X$}(0) &=&  0	\\
	\mbox{\boldmath$X$}(s^*) &=&  \mbox{\boldmath$I$}
\end{eqnarray}

\noindent where {\boldmath$I$} is the identity matrix. The vanishing
boundary condition at the observer ensures that the observer receives
all the neighboring lightrays.  The unit boundary condition at the
source ensures that the two Jacobi fields that we are solving for are
orthonormal and aligned with the basis $(e_1^a,e_2^a)$ at the source
location. [It is worthwhile noticing that we could have, just as well,
imposed boundary conditions that would allow us to obtain the basis
$(\widetilde{M}_1^a, \widetilde{M}_2^a)$. These would require that
{\boldmath$X$} be vanishing and $D\mbox{\boldmath$X$}$ be the identity
at $s=0$. By Eq.~(\ref{consist1}) and Eq.~(\ref{consist2}), both procedures are
equivalent for the purposes of finding the magnification matrix. ]

At all points but $s=L$, Eq.~(\ref{eqn1}) reduces to
$D^2\mbox{\boldmath$X$}=0$, thus the solution has the form of two linear
functions of $s$ at both sides of the lens plane. We denote these
functions as $\mbox{\boldmath$X$}_O$ and $\mbox{\boldmath$X$}_S$ for the
observer's side and the source's side, respectively:

\begin{equation}
	\mbox{\boldmath$X$}
   =
	\left\{\begin{array}{ll}
	\mbox{\boldmath$X$}_O \equiv s \mbox{\boldmath$K$}	&	
	L<s\leq 0						\\
	\mbox{\boldmath$X$}_S \equiv (s-s^*) \mbox{\boldmath$N$}
				+\mbox{\boldmath$I$}\hspace{1cm}&
	s^*\leq s\leq L	
		\end{array}
	\right.
\label{startx}
\end{equation}

\noindent where {\boldmath$K$} and {\boldmath$N$} are two constant
matrices which are determined by matching conditions on the lens plane.
The matching conditions, according to Eqs.~(\ref{eqn1}) and (\ref{Tdef}), are
continuity of {\boldmath$X$} and a jump in the first derivative of
{\boldmath$X$}, namely

\begin{eqnarray}
   \mbox{\boldmath$X$}_O(L) &=& \mbox{\boldmath$X$}_S(L),   \label{bc1}\\
  D\mbox{\boldmath$X$}_O(L) &=& D\mbox{\boldmath$X$}_S(L) 
				-\mbox{\boldmath$TX$}_S(L). \label{bc2}
\end{eqnarray}

\noindent This allows us to obtain {\boldmath$K$} from the curvature
matrix {\boldmath$T$} as

\begin{equation}
     \mbox{\boldmath$K$} 
   = 
	\Big( s^*\mbox{\boldmath$I$} - L(L-s^*)\mbox{\boldmath$T$}
	\Big)^{-1}
\label{K}
\end{equation}

\noindent and {\boldmath$N$} from {\boldmath$K$} and {\boldmath$T$} as

\begin{equation}
     \mbox{\boldmath$N$} 
   = 
	 \mbox{\boldmath$K$} +L \mbox{\boldmath$TK$}
\label{NfromKT}
\end{equation}

\noindent Our goal is, however, to find {\boldmath$K$} in terms of the
optical scalars, so we need to express {\boldmath$T$} in terms of
{\boldmath$P$}.  This can be done very quickly here because of the
thin-lens regime that we are imposing.  From Eq.~(\ref{eqn3}) we obtain
$\mbox{\boldmath$K$}= -s\mbox{\boldmath$P$}_O \mbox{\boldmath$K$}$, or

\begin{equation}
	\mbox{\boldmath$P$}_O = -\frac{\mbox{\boldmath$I$}}{s}
\end{equation}

\noindent on the observer's side or the lens plane, as it should be since
on the observer's side the lightcone is the same as in flat spacetime. 
On the source's side, however, we obtain 

\begin{equation}\label{NfromP}
     \mbox{\boldmath$N$} 
   = 
	-\mbox{\boldmath$P$}_S
	\Big( (s-s^*) \mbox{\boldmath$N$}
		      +\mbox{\boldmath$I$}
	\Big),
\end{equation}

\noindent which is valid at all values of $s$ on the source's side. In
particular, we evaluate it at $s=L$ and use Eq.~(\ref{bc1}) on the
right-hand-side and Eq.~(\ref{NfromKT}) on the left-hand-side, obtaining

\begin{equation}
	   \mbox{\boldmath$K$} 
	+L \mbox{\boldmath$TK$}
   =
	-L\mbox{\boldmath$P$}_S(L)
	 \mbox{\boldmath$K$},
\end{equation}
		
\noindent or

\begin{equation}
	   \mbox{\boldmath$P$}_S(L) 
   =
	-\frac{\mbox{\boldmath$I$}}{L}
	 -\mbox{\boldmath$T$}.
\label{Psol}
\end{equation}

\noindent This means that the curvature is related to the values of the
optical scalars at the lens plane (on the source's side) by

\begin{equation}
	\rho_S(L) = -\left(\frac{1}{L}+\widetilde{\Phi}_{00}\right),
	\hspace{1cm}
	\sigma_S(L) = - |\widetilde{\Psi}_0|e^{i\varphi}.
\end{equation}

\noindent Substituting Eq.~(\ref{Psol}) into Eq.~(\ref{K}) we obtain

\begin{equation}
     \mbox{\boldmath$K$} 
   = 
	\frac{1}{L}\Big(\mbox{\boldmath$I$} + (L-s^*)\mbox{\boldmath$P$}_S(L)
	\Big)^{-1}
   =
	\frac{1}{L\Delta}
	\left(\begin{array}{cc}
		1 + (L-s^*)\rho_L \hspace{.3cm}  &  -(L-s^*)\bar\sigma_L	\\
		   -(L-s^*)\sigma_L &   1 + (L-s^*)\rho_L 
			\end{array}
		\right)
\label{Ksol}
\end{equation}

\noindent with 

\begin{equation}
	\Delta \equiv (1+(L-s^*)\rho_L)^2
		    -(L-s^*)^2\sigma_L\bar\sigma_L,
	\hspace{1cm}
	\rho_L\equiv \rho_S(L),
	\hspace{1cm}
	\sigma_L\equiv \sigma_S(L).
\end{equation}

\noindent For completeness, we display the form of the shear and divergence in
the source's side of the lens, which can be obtained directly from
Eq.~(\ref{eqn2}) by setting $\mbox{\boldmath$Q$}=0$ (or, less directly, from Eqs.~(\ref{NfromKT}),  (\ref{NfromP}) and (\ref{Ksol}) we can
solve for $\mbox{\boldmath$P$}_S$):

\begin{eqnarray}
	\rho_S(s) &=& \frac{(\rho_L^2-\sigma_L\bar{\sigma}_L)
			   \Big(\rho_L-(s-L)(\rho_L^2-\sigma_L\bar{\sigma}_L)
			\Big)}
			   {\Big(\rho_L-(s-L)(\rho_L^2-\sigma_L\bar{\sigma}_L)
			\Big)^2
			    -\sigma_L\bar{\sigma}_L}			\\
	\sigma_S(s) &=& \frac{(\rho_L^2-\sigma_L\bar{\sigma}_L)
			    \sigma_L}
			   {\Big(\rho_L-(s-L)(\rho_L^2-\sigma_L\bar{\sigma}_L)
			\Big)^2
			    -\sigma_L\bar{\sigma}_L}
\end{eqnarray}

We can now read off the complex components of the complex
connecting vectors $M^a$ and use them to obtain the real connecting
vectors $M_1^a$ and $M_2^a$.  We have 

\begin{eqnarray}
     M_1^a
   &=&
	 \frac{(\xi+\bar\xi+\eta+\bar\eta)}{2}e_1^a
	+ \frac{(\eta-\bar\eta-\xi+\bar\xi)}{2i}e_2^a	\\
     M_2^a
   &=&
	 \frac{(\xi-\bar\xi+\eta-\bar\eta)}{2i}e_1^a
	+ \frac{(\xi+\bar\xi-\eta-\bar\eta)}{2}e_2^a	
\end{eqnarray}

\noindent which, with Eq.~(\ref{Ksol}) and Eq,~(\ref{K}), yields

\begin{eqnarray}
     M_1^a
   &=&
	\frac{s}{L\Delta}(1+(L-s^*)\rho_L-(L-s^*)|\sigma_L|\cos\varphi)
	     e_1^a
	- \frac{s}{L\Delta}((L-s^*)|\sigma_L|\sin\varphi)
	      e_2^a	\\
     M_2^a
   &=&
   	- \frac{s}{L\Delta}((L-s^*)|\sigma_L|\sin\varphi)
	      e_1^a
	+\frac{s}{L\Delta}(1+(L-s^*)\rho_L+(L-s^*)|\sigma_L|\cos\varphi)
	     e_2^a.		
\end{eqnarray}

\noindent Consequently, by Eq.~(\ref{jay}), the components of the matrix
\mbox{\boldmath$J$} on the observer's side are 

\begin{equation}
\mbox{\boldmath $J$}
   =
	\frac{s}{L\Delta}
	\left(\begin{array}{cc}
		1 + (L-s^*)\rho_L -(L-s^*)|\sigma_L|\cos\varphi  &  
		-(L-s^*)|\sigma_L|\sin\varphi			\\
	&\\
		-(L-s^*)|\sigma_L|\sin\varphi	 		&   
		1 + (L-s^*)\rho_L +(L-s^*)|\sigma_L|\cos\varphi
	      \end{array}
	\right).
\end{equation}	
By Eq.~(\ref{consist2}), our magnification matrix  
${}^*\!\!\widetilde{\mbox{\boldmath $J$}}^{-1}$, which maps the source's
shape into0 the image's shape, is thus

\begin{equation}\label{Jstar-1}
{}^*\!\!\widetilde{\mbox{\boldmath $J$}}^{-1}
   =
	\frac{1}{L\Delta}
	\left(\begin{array}{cc}
		1 + (L-s^*)\rho_L -(L-s^*)|\sigma_L|\cos\varphi  &  
		-(L-s^*)|\sigma_L|\sin\varphi			\\
	&\\
		-(L-s^*)|\sigma_L|\sin\varphi	 		&   
		1 + (L-s^*)\rho_L +(L-s^*)|\sigma_L|\cos\varphi
	      \end{array}
	\right).
\end{equation}

\noindent To obtain the Jacobian matrix \mbox{\boldmath $A$} of the
thin-lens approach, we use Eq.~(\ref{AfromJ}), which takes care of the
scaling of the source vector. Inverting Eq.~(\ref{Jstar-1}) and dividing by
$s^*$ we obtain

\begin{equation}
\mbox{\boldmath $A$}
   =
	\frac{L}{s^*}
	\left(\begin{array}{cc}
		1 + (L-s^*)\rho_L +(L-s^*)|\sigma_L|\cos\varphi  &  
		(L-s^*)|\sigma_L|\sin\varphi			\\
	&\\
		(L-s^*)|\sigma_L|\sin\varphi	 		&   
		1 + (L-s^*)\rho_L -(L-s^*)|\sigma_L|\cos\varphi
	      \end{array}
	\right).
\end{equation}

\noindent By comparison with Eq.~(\ref{Astandard}),  we can make the
following identifications

\begin{eqnarray}
	\kappa &=& \frac{(s^*-L)}{s^*}(1+L\rho_L),\\
	\gamma_1 &=&\frac{L}{s^*}(s^*-L)|\sigma_L|\cos\varphi,\\
	\gamma_2 &=&\frac{L}{s^*}(s^*-L)|\sigma_L|\sin\varphi.
	\label{thinlensparameters}
\end{eqnarray}

\noindent In the weak-field, thin-lens approach to lensing, $\kappa$ and 
$\gamma\equiv\sqrt{\gamma_1^2+\gamma_2^2}$ are referred to as the
convergence and shear.  We can now see how they relate to the values of the
optical scalars of the lightcone -- the convergence and shear--, evaluated
at the lens plane on the source's side.  We can verify, by inspection, that
$\kappa, \gamma_1$ and $\gamma_2$ vanish in the case of flat space, where
$\rho_L=-1/L$ and $\sigma_L=0$, which means that there is no distortion in
the absence of a lens (in the thin-lens regime, it makes sense to define
distortion away from the distortion associated with flat space, which is,
naturally, the convergence of the lightcone).  

The generic case in which the curvature along a lightray is continuous also
offers a relationship between image distortion and the optical scalars
$\rho$ and $\sigma$, certainly not in terms of the values of the optical
scalars at particular points. We phrase this relationship in the context of
the shape parameters, rather than the magnification matrix.

\section{Generic non-perturbative approach to image distortion}


In this section we generalize the results of the previous section to the
case of generic spacetimes.  Our goal is to find a direct relationship
between the optical scalars and the distortion of the cross-section of
the pencil of rays connecting the observer to an elliptical source.
Such a relationship  will be interpreted as the direct effect of
spacetime curvature in the distortion of small elliptical images. 

That such a relationship exists can be seen from the following. By
Eqs.~(\ref{distR})-(\ref{distdelta}), the distortion parameters ${\frak
D}_{\Omega},{\frak D}_{\cal R}$ and  ${\frak D}_{\delta}$ are expressed in
terms of the scalar products of the Jacobi basis $(M_1^a, M_2^a)$.  The
Jacobi basis vectors $(M_1^a, M_2^a)$ along a fixed lightray need not be
derived from the lightcone, but can be obtained directly from the optical
scalars by means of Eq.~(\ref{eqn3}), as we showed in the previous section
for a special case.  Therefore, the distortion parameters can be thought of
as functionals of the optical scalars.  Here we present one way of
obtaining an explicit relationship  between thedistortion parameters and
the optical scalars.  The relationship  that we obtain is a system of
first-order ordinary differential equations for the evolution of the shape
parameters along a fixed lightray, where the optical scalars enter as known
sources.  This system of equations can be thought of as an extension of the
focusing equation for the determinant of the Jacobian matrix in
lensing~\cite{EFS}. Significant differences are, however, that the
equations are of first order (instead of second order) and that all three
parameters are evolved, not just the area of the beam. 

To obtain this result we start by taking a derivative of the shape
parameters with respect to the affine parameter $s$ and evaluating the
derivative at the value $s^*$, at which the Jacobi basis is {\it
orthonormal\/}.  This can be regarded as an infinitesimal distortion, in
a sense to be made specific. We have, from Eq.~(\ref{area}), for the
infinitesimal distortion of the area: 

\begin{equation}
\left.\frac{d{\cal A}^2}{ds} \right|_{s^*}
   =
	-{}^*\!\!\!{\cal A}^2\left.\frac{d}{ds}\big(M_1\!\cdot M_1
		+M_2\!\cdot M_2\big)\right|_{s^*}
\label{adotmed}
\end{equation}

The infinitesimal distortion of the semiaxes ratio can be obtained from
Eq.~(\ref{ratio}) by

\begin{equation}
\left.\frac{d{\cal R}^2}{ds} \right|_{s^*}
   =
	\left.
	\left(
	\frac{\partial{\cal R}^2 }{\partial a} \frac{da}{ds}
	+\frac{\partial{\cal R}^2}{\partial b}  \frac{db}{ds}
	+\frac{\partial{\cal R}^2}{\partial c}  \frac{dc}{ds}
	\right)
	\right|_{s^*},
\label{rdotstart}
\end{equation}

\noindent where $(a,b,c)$ are the functions of $s$ defined by
Eqs.~(\ref{a})-(\ref{c}). From Eqs.~(\ref{a}), (\ref{b}) and (\ref{c}) we
can see that 

\begin{eqnarray}
	a(s^*)&=&-{}^*\!{\cal R}^2	,\label{a*}\\
	b(s^*)&=&{}^*\!{\cal R}^2-1	,\label{b*}\\
	c(s^*)&=&0.
\label{c*}			
\end{eqnarray}

\noindent By calculating the partial derivatives of Eq.~(\ref{ratio}) and
using Eqs.~(\ref{a*})-(\ref{c*}) we obtain

\begin{eqnarray}
\left.\frac{\partial{\cal R}^2 }{\partial a}\right|_{s^*}
   &=& 	
	{}^*\!{\cal R}^2 -1			,\label{ra*}\\
\left.\frac{\partial{\cal R}^2 }{\partial b}\right|_{s^*}
   &=&
	{}^*\!{\cal R}^2				,\label{rb*}\\
\left.\frac{\partial{\cal R}^2 }{\partial c}\right|_{s^*}
   &=&
	0.
\label{rc*}
\end{eqnarray}

\noindent Taking an $s-$derivative of Eqs.~(\ref{a})-(\ref{c}), evaluating
it at $s^*$ and using the resulting expression, together with
Eqs.~(\ref{ra*})-(\ref{rc*}), in Eq.~(\ref{rdotstart}) yields

\begin{equation}
\left.\frac{d{\cal R}^2}{ds}\right|_{s^*}
   =
	{}^*\!{\cal R}^2
	\left.
	\left(
	\cos(2\delta)\frac{d}{ds}
		\big(M_2\!\cdot M_2
		     -M_1\!\cdot M_1\big)
	-2\sin(2\delta)\frac{d}{ds}
		M_1\!\cdot M_2		     
	\right)
	\right|_{s^*}.
\label{rdotmed}
\end{equation}

\noindent The infinitesimal distortion of the orientation parameter is more
complicated but can be obtained in a straightforward manner via a similar
procedure:

\begin{equation}
\left.\frac{d\delta}{ds} \right|_{s^*}
   =
	\left.
	\frac{({}^*\!{\cal R}^2+1)}
	     {({}^*\!{\cal R}^2-1)}
	\left(\frac{\sin(2\delta)}{4}
	\frac{d}{ds}\big( M_1\!\cdot M_1
		         -M_2\!\cdot M_2\big) 
	-\frac{\cos(2\delta)}{2}
	\frac{d}{ds}M_1\!\cdot M_2
	\right)\right|_{s^*}.
\label{deltadotmed}
\end{equation}

Notice that Eqs.~(\ref{adotmed}), (\ref{rdotmed}) and (\ref{deltadotmed})
depend on the derivatives of the scalar products between our Jacobi basis
vectors.  In order to eliminate these in terms of the optical scalars we
find first an expression for the optical scalars in terms of them.  These
can be obtained directly from their definitions, Eqs.~(\ref{rhodef}) and
(\ref{sigmadef}), using Eq.~(\ref{ma}) and Eqs.~(\ref{e1})-(\ref{e2}). We
have

\begin{equation}
	m^a = \frac{\alpha}{\sqrt{2}} e^{i(\lambda+\nu)}M_1^a
		+\frac{\beta}{\sqrt{2}} e^{i\lambda}M_2^a
\end{equation}

\noindent so

\begin{eqnarray}
	\rho
   &=&  \frac12
	\Big( \alpha^2 M_1^aM_1^b
		+\beta^2   M_2^aM_2^b
		+2\alpha\beta \cos\nu (M_1^aM_2^b+M_1^bM_2^a)
	\Big)
	\nabla_a\ell_b					\label{rhomed}\\	
	\sigma 
   &=& \frac12
	\Big(    \alpha^2 e^{2i(\lambda+\nu)} M_1^aM_1^b
		+\beta^2  e^{2i\lambda} M_2^aM_2^b
		+\alpha\beta e^{i(2\lambda+\nu)} (M_1^aM_2^b+M_1^bM_2^a)
	\Big)
	\nabla_a\ell_b					\label{sigmamed}					
\end{eqnarray}

\noindent where we have used the fact that $\nabla_b\ell_a=\nabla_a\ell_b$.
In order to put Eqs.~(\ref{rhomed})-(\ref{sigmamed}) in a convenient form,
we use the fact that that the connecting vectors $M_1^a, M_2^a$ are
Lie-dragged by the geodesic vector $\ell^a$, namely, Eq.~(\ref{lie})
applied to both $M_1^a$ and $M_2^a$:

\begin{eqnarray}
	M_1^a\nabla_a\ell^b &= &\ell^a\nabla_a M_1^b,	\\
	M_2^a\nabla_a\ell^b &= &\ell^a\nabla_a M_2^b.
\end{eqnarray}

\noindent This allows us to trade the covariant derivatives of $\ell_a$ for
ordinary derivatives of the scalar products between the two connecting
vectors:

\begin{eqnarray}
	\rho &=& \frac14\left(
		 \alpha^2 \frac{d}{ds}(M_1\!\cdot\!M_1)
		+\beta^2 \frac{d}{ds}(M_2\!\cdot\!M_2)
		+2\alpha\beta\cos\nu \frac{d}{ds}(M_1\!\cdot\!M_2)
			\right)	,		\label{rhoprelim}\\
	\sigma &=& \frac{e^{2i\lambda}}{4} \left(
		   \alpha^2 e^{2i\nu}\frac{d}{ds}(M_1\!\cdot\!M_1)
		  +\beta^2 \frac{d}{ds}(M_2\!\cdot\!M_2)
		+2\alpha\beta e^{i\nu} \frac{d}{ds}(M_1\!\cdot\!M_2)
					   \right). \label{simgaprelim}
\end{eqnarray}

\noindent For completeness, we display the final expressions for the
optical scalars in terms of the connecting vectors, although
Eqs.~(\ref{rhoprelim}) and (\ref{simgaprelim}) are sufficient for the
purposes of this section. Substituting Eqs.~(\ref{alpha}), (\ref{beta}) and
(\ref{nu}) in Eq.~(\ref{rhoprelim}), the expression for $\rho$ becomes

\begin{equation}
	\rho = -\frac14 \frac{d}{ds}
			\log(M_1\!\cdot M_1 M_2\!\cdot M_2
			- (M_1\!\cdot M_2)^2).
\label{rhofinal}
\end{equation}	

\noindent Using Eqs.~(\ref{alpha}), (\ref{beta}) and (\ref{nu}) in
Eq.~(\ref{simgaprelim}), we obtain an expression for the magnitude of
$\sigma$:

\begin{eqnarray}
	\sigma\bar\sigma 
   &=&
	\frac{(M_1\!\cdot\!M_1 M_2\!\cdot\!M_2)^2}
	     {4 (M_1\!\cdot M_1 M_2\!\cdot M_2
			- (M_1\!\cdot M_2)^2)^2} \nonumber\\
   && \times
	\left( \left(\frac{d}{ds}\log\!
			\left(\!\frac{M_1\!\cdot\!M_1}
				   {M_2\!\cdot\!M_2}
			\right)\!\!
		\right)^2\!
	      + \frac{4(M_1\!\cdot\!M_2)^2}
		      {M_1\!\cdot\!M_1 M_2\!\cdot\!M_2}
		\frac{d}{ds}\log\!
			\left(\!\frac{M_1\!\cdot\!M_2}
				   {M_1\!\cdot\!M_1}
			\right)
		 \frac{d}{ds}\log\!
			\left(\!\frac{M_1\!\cdot\!M_2}
				   {M_2\!\cdot\!M_2}
			\right)
	\right).
\label{sigmafinala}
\end{eqnarray}

\noindent In order to obtain an expression for the phase $\varphi$ of
the shear ($\sigma\equiv|\sigma|e^{i\varphi}$), we first calculate the
phase of $\hat{\sigma}\equiv |\sigma|e^{i\hat{\varphi}}$, substituting
Eqs.~(\ref{alpha}), (\ref{beta}) and (\ref{nu}) into Eq.~(\ref{simgaprelim}):

\begin{equation}
	\tan\hat{\varphi}
   =
	\frac{ 2 M_1\!\cdot\!M_2 \sqrt{A_s^2/2}
		\frac{d}{ds}\log \left(\frac{M_1\!\cdot\!M_2}
				   	    {M_1\!\cdot\!M_1}
				\right)				}
	     {M_1\!\cdot\!M_1 M_2\!\cdot\!M_2
		\frac{d}{ds}\log\left( \frac{M_1\!\cdot\!M_1}
				   	   {M_2\!\cdot\!M_2}
				\right)
		+ 2 (M_1\!\cdot\!M_2)^2
		\frac{d}{ds}\log \left(\frac{M_1\!\cdot\!M_2}
				   	    {M_1\!\cdot\!M_1}	
				\right)				}.
\end{equation}

\noindent The phase of the shear is thus

\begin{equation}
	\phi = 2\lambda + \hat{\varphi},
\label{sigmafinalb}
\end{equation}

\noindent where $\lambda$ is determined by Eq.~(\ref{lambdadot}). The set of
equations Eqs.~(\ref{rhofinal}), (\ref{sigmafinala}) and (\ref{sigmafinalb})
is the equivalent of Eq.~(\ref{eqn3}) of the previous Section. For our
purposes, however, we use the preliminary form of the equations given by
Eqs.~(\ref{rhoprelim}) and (\ref{simgaprelim}).  We evaluate both
Eq.~(\ref{rhoprelim}) and Eq.~(\ref{simgaprelim}) at $s^*$, using
Eq.~(\ref{starredvalues}) for $\alpha, \beta, \nu$ and $\lambda$.  We obtain

\begin{eqnarray}
	\rho_* &=& \frac14\left.\frac{d}{ds}\big(
		  M_1\!\cdot\!M_1
		+ M_2\!\cdot\!M_2\big)
			\right|_{s^*}	,		\label{rho*}\\
	\sigma_* &=& \frac14 \left.\frac{d}{ds}\big(
		   M_1\!\cdot\!M_1
		  -M_2\!\cdot\!M_2
		+2iM_1\!\cdot\!M_2		\big)
					   \right|_{s^*}. \label{sigma*}
\end{eqnarray}

\noindent Subtituting Eq.~(\ref{rho*}) and Eq.~(\ref{sigma*}) into 
Eq.~(\ref{adotmed}),Eq.~(\ref{rdotmed}) and Eq.~(\ref{deltadotmed}), we obtain

\begin{eqnarray}
\left.\frac{d{\cal A}^2}{ds} \right|_{s^*}
   &=&
	-4{}^*\!\!\rho\;	{}^*\!\!\!{\cal A}^2		\\
\left.\frac{d{\cal R}^2}{ds}\right|_{s^*}
   &=&
	-2
	\Big(
	 \cos(2{}^*\!\delta)
		({}^*\!\!\sigma+{}^*\!\!\bar{\sigma})
	-i\sin(2{}^*\!\delta)
		({}^*\!\!\sigma-{}^*\!\!\bar{\sigma})		     
	\Big){}^*\!{\cal R}^2			\\
\left.\frac{d\delta}{ds} \right|_{s^*}
   &=&
	\frac{({}^*\!{\cal R}^2+1)}
	     {2({}^*\!{\cal R}^2-1)}
	\Big(\sin(2{}^*\!\delta)
	 ({}^*\!\!\sigma+{}^*\!\!\bar\sigma) 
	+i\cos(2{}^*\!\delta)
	({}^*\!\!\sigma-{}^*\!\!\bar\sigma)\Big).
\end{eqnarray}

\noindent Since these equations hold for arbitrary fixed value of $s^*$,
we can drop the stars and consider them as differential equations for
the shape parameters

\begin{eqnarray}
\frac{d{\cal A}^2}{ds}
   &=&
	-4\rho\;	{\cal A}^2		\label{adotfinal}\\
\frac{d{\cal R}^2}{ds}
   &=&
	-2
	\Big(
	 \cos(2\delta)
		(\sigma+\bar\sigma)
	-i\sin(2\delta)
		(\sigma-\bar\sigma)		     
	\Big){\cal R}^2			\label{rdotfinal}\\
\frac{d\delta}{ds} 
   &=&
	\frac{({\cal R}^2+1)}
	     {2({\cal R}^2-1)}
	\Big(\sin(2\delta)
	 (\sigma+\bar\sigma) 
	+i\cos(2\delta)
	(\sigma-\bar\sigma)\Big).		\label{deltadotfinal}
\end{eqnarray}

\noindent Notice that, as a consequence of Eq.~(\ref{adotfinal}), the solid
angle $\Omega\equiv{\cal A}/s^2$ satisfies the differential
equation

\begin{equation}
\frac{d\Omega^2}{ds}
   =
	-4\left(\rho+\frac{1}{s}\right)	\Omega^2,
\label{ahatdotfinal}
\end{equation}

\noindent which shows that in flat space, where $\rho=-1/s$, there is no
change in the solid angle of the image $\Omega_I$ as compared to the
solid angle of the source ${}^*\Omega$, since there is no change in the
solid angle of the pencil of rays $\Omega(s)$. We consider alternatively
Eq.~(\ref{adotfinal}) or Eq.~(\ref{ahatdotfinal}) as equally relevant in order
to refer to infinitesimal area distortion or infinitesimal solid angle
distortion. However, it is the infinitesimal solid angle distortion
that is relevant to total distortion, as we see immediately below.   

Equations~(\ref{adotfinal}) (or equivalently (\ref{ahatdotfinal})),
(\ref{rdotfinal}) and (\ref{deltadotfinal}) constitute our desired
direct relationship between image distortion and the optical scalars. 
In these equations, the optical scalars $\rho$ and $\sigma$ are thought
of as given, since they are obtained, along a fixed null ray, from
Eq.~(\ref{eqn2}) where the spacetime curvature acts as the source. As
expected, the convergence $\rho$ (also referred to as divergence and
expansion elsewhere) enters only in the equation for the area
distortion.  Notice, however, the role of both the real and imaginary
parts of the shear $\sigma$ in both the orientation and semiaxes ratio
distortions.  Notice, as well, that Eq.~(\ref{deltadotfinal}) breaks
down when the ratio of the semiaxes takes the value 1, which reflects
the fact that the orientation is ill defined at the points where the
shape of the pencil of rays is circular. 

The finite distortion parameters ${\frak D}_{\Omega},{\frak D}_{\cal R}$
and ${\frak D}_{\delta}$ are obtained by direct integration of the
infinitesimal distortions:

\begin{eqnarray}
	{\frak D}_{\Omega}
   &=&\frac{1}{{}^*\Omega^2}
	\int_{s^*}^0 \frac{d\Omega^2}{ds} ds
   = \left(\frac{\Omega_I}{{}^*\Omega}\right)^2 -1\\
	{\frak D}_{\cal R}
   &=&\frac{1}{{}^*\!{\cal R}^2}
	\int_{s^*}^0 \frac{d{\cal R}^2}{ds} ds
= \left(\frac{{\cal R}_I}{{}^*\!{\cal R}}\right)^2 -1\\
	{\frak D}_{\delta}
   &=&
	\int_{s^*}^0 \frac{d\delta}{ds}ds
   =   \delta_I-{}^*\!\delta.
\end{eqnarray}

\noindent As discussed in~\cite{FKNI}, we say that there has been image 
distortion if any of the three distortion parameters is non-zero. As we
noted, in flat space, $\rho = -1/s$, so there is no  change in the solid
angle in Eq.~(\ref{ahatdotfinal}) and the distortion  parameter ${\frak
D}_{\Omega}$ is zero. Likewise, $\sigma$ is zero in flat space, so ${\frak
D}_{\cal R}$ and ${\frak D}_\delta$ are also zero by virtue of 
Eqs.~(\ref{rdotfinal}) and (\ref{deltadotfinal}).

In order to illustrate the significance of Eqs. (\ref{adotfinal}),
(\ref{rdotfinal}) and (\ref{deltadotfinal}), in the following subsection we
specialize to the case of a Schwarzschild spacetime.

\subsection*{The Schwarzschild case}


Consider a Schwarzschild spacetime in standard Schwarzschild coordinates

\begin{equation}\label{line}
    ds^2=\left(1-\frac{2m}{r}\right)dt^2
        -\left(1-\frac{2m}{r}\right)^{-1}dr^2 
	-r^2(d\theta^2+\sin^2\!\theta d\phi^2).
\end{equation}

\noindent The deflector and the observer define a coordinate line that
is fixed for the lensing problem, referred to as the optical axis.  For
simplicity, we choose this to be the $z-$axis.  Because the spacetime is
spherically symmetric, the lensing problem (defined essentially by the
past lightcone of the observer located on the $z-$axis) has axial
symmetry around the optical axis.  Every geodesic emitted  ``backwards''
in time from the observer remains on a coordinate plane with a fixed
value of $\phi$. Therefore, there is a connecting vector, $M_1^a$, that
lies on the plane $\phi=constant$ (i.e, has no component along
$\partial/\partial\phi$) and there is a second connecting vector,
$M_2^a$,  that joins geodesics in adjacent $\phi=constant$ planes (i.e.,
can be taken to have only a component along $\partial/\partial \phi$).  
With the metric, Eq.~(\ref{line}), these two connecting vectors are
orthogonal everywhere on the lightcone.  In terms of these connecting
vectors we have that 

\begin{equation}
	\sigma = \frac14 
		\frac{d}{ds}
		\log\left( \frac{M_1\!\cdot\!M_1}
				{M_2\!\cdot\!M_2}
		    \right),
\label{schwsigma}
\end{equation}

\noindent which follows from Eq.~(\ref{simgaprelim}) by using the facts
that $\nu=constant=\pi/2$ and $\lambda=constant=0$ (by choice), both of
which are consequences of $M_1\!\cdot\!M_2 =0$ applied to Eqs.~(\ref{nu})
and (\ref{lambdadot}).   Thus the shear is real (up to a constant phase
that has been chosen as zero).   

If the shear is real, the differential equation for the
orientation parameter admits the solution $\delta(s)=0$.  Thus, if
$\delta$ starts out with the value zero, it remains zero along the given
null ray. With $\delta=0$, the equation for the ratio of the semiaxes,
Eq.~(\ref{rdotfinal}), reads 

\begin{equation}
\frac{d{\cal R}^2}{ds}
   =
	-4\sigma{\cal R}^2
\end{equation}

\noindent with solution

\begin{equation}
	{\cal R}^2 = {}^*\!{\cal R}^2\exp\left(-4\int_{s^*}^s\sigma\right)
\end{equation}

\noindent If one uses Eq.~(\ref{schwsigma}), it can be seen that this
agrees with the direct calculation of the ratio carried out in our
companion paper~\cite{FKNI}.  

On the other hand, if $\delta$ is not vanishing at the start, then it does
not remain constant, by Eq.~(\ref{deltadotfinal}), and we have a system of
coupled differential equations for ${\cal R}$ and $\delta$.   This reflects
the fact that the images of sources that respect the axial symmetry of the
lensing  problem also respect such symmetry.  However, the images of
sources that do not have axial symmetry around the optical axis will be
rotated, as described in our companion paper~\cite{FKNI}.

\section{Concluding remarks}


We have attempted to provide a unified description of image distortion in a
non-perturbative (exact) manner within the framework of general relativity.
We have tried to formalize our description in a manner as close as possible
to the standard (weak-field thin-lens) treatment with the hope that it will
be useful in deepening the current understanding of the gravitational
lensing phenomenon.  

We have related the convergence and shear used in
standard lensing to the actual values of the geometrically defined optical
scalars (convergence and shear) at the lens plane.  

The roles of our non-perturbative distortion parameters can be compared
with those of the convergence and shear from the thin-lens approximation.
Both descriptions of image distortion are defined such that they vanish in
flat space, so that a non-zero value indicates a deviation from an
``unlensed'' image. The $\kappa$ in the thin-lens approximation is
analogous to ${\frak D}_\Omega$ in that they both give an indication of an
overall change in angular size. The $\gamma=\sqrt{\gamma_1^2+\gamma_2^2}$
in Eqs.~(\ref{thinlensparameters}) indicates stretching and squeezing of
the image along perpendicular axes, as ${\frak D}_{\cal R}$ does.  Lastly,
the ratio $\gamma_2/\gamma_1$ gives the orientation of the axes of the
image, as ${\frak D}_\delta$ does. The analogy is not perfect, but the
issue is that there are as many parameters to completely describe image
distortion in one case as in the other.  

Clearly, image distortion is ultimately governed by the geodesic deviation
equations for the connecting vectors of the pencil of rays between the
source and the observer.  This is a system of second-order equations for
four variables (the two components of two linearly independent connecting
vector fields which span the space of solutions, or, equivalently, the four
components of the 2-dimensional linear map $D_i^j$ that maps the initial
values of a connecting vector $\dot{Z}_j(0)$ into its values at a point $s$
via $Z_i(s)=D_i^j\dot{Z}_j(0)$).  Still, the distortion of elliptical
images has only three relevant parameters. Defining the three distortion
parameters and using their evolution equations in terms of the optical
scalars, (\ref{rdotfinal}), (\ref{deltadotfinal}) and (\ref{ahatdotfinal}),
allows us two advantages:  1) the problem is a system of equations for the
right number of relevant quantities, and 2) the system is of first order,
rather than second. 

In the end, defining the distortion parameters allows us to reduce, in a
way, the geodesic deviation equations down to the specific problem of
elliptical images. We feel that they may become useful in studies of weak
lensing with multiple lens planes, cosmic shear~\cite{Tyson} and distortion
of high-redshift sources~\cite{Blandford-Jaroszynski}.  In particular,
meaningful observables can actually be defined out of the shape and 
distortion parameters.  Work on this issue is in progress and will be reported
elsewhere.

\acknowledgments

This work was supported by the NSF under grants No. PHY 98-03301, PHY
92-05109 and PHY 97-22049. We are indebted to Volker Perlick for kindly
pointing out to us Ref.~\cite{Panov}, and to J\"{u}rgen Ehlers for
stimulating conversation.



\newpage 
\begin{figure}
\centerline{\psfig{figure=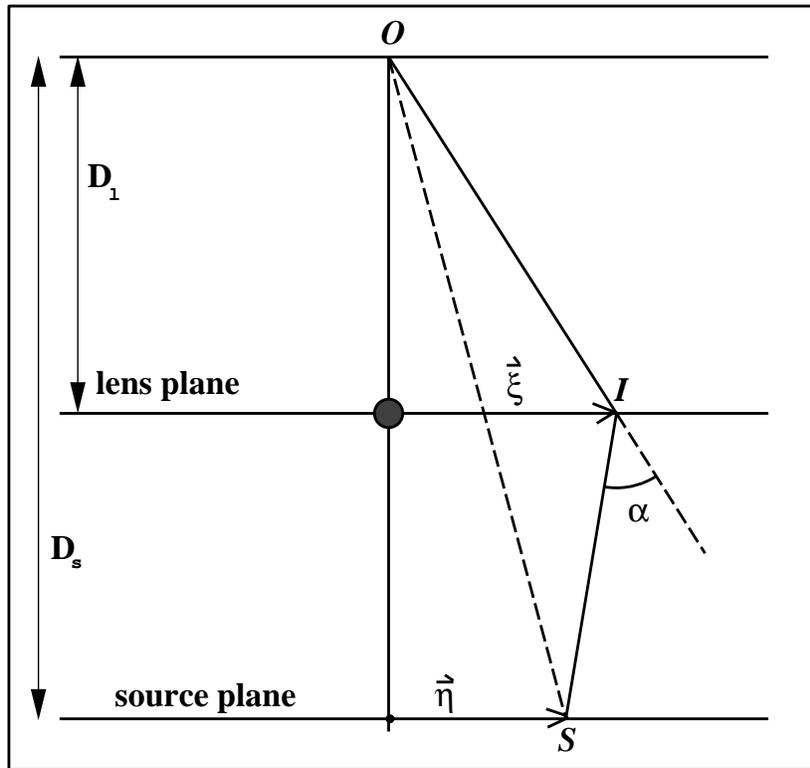,height=4in,angle=0} }

\caption{The lensing problem in standard gravitational lensing.  The
point $O$ represents the observer.  The point $S$ represents the source.
The lens mapping can be interpreted as a map that takes the point $I$ in
the lens plane into the point $S$ in the source plane.}

\label{fig:distIIa}
\end{figure}

\newpage
\begin{figure}
\centerline{\psfig{figure=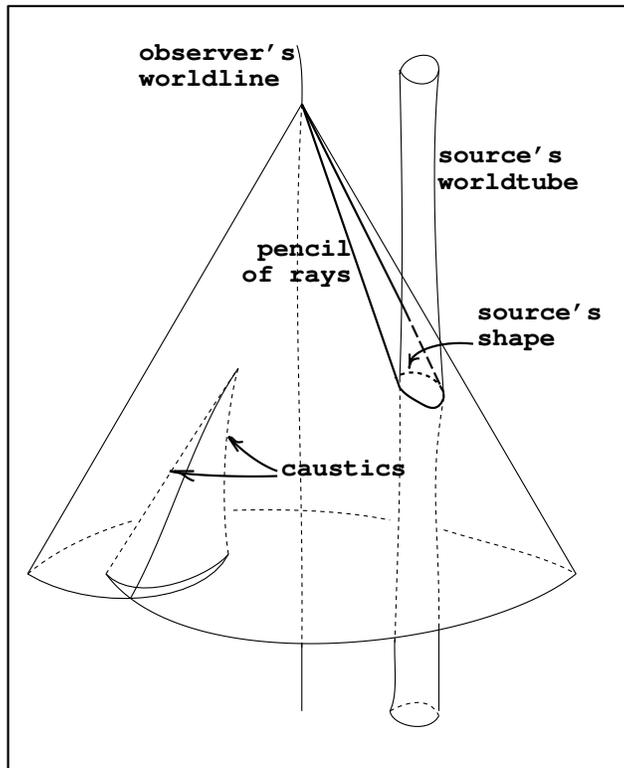,height=4in,angle=0}}

\caption{Diagram of image distortion.  The source's worltube intersects the
observer's past lightcone in a region free of caustics.  The source's
visible shape is defined by the intersection.  The pencil of rays between
the source's shape and the observer carries the shape of the source into
the shape of the image, on the observer's celestial sphere. The pencil of
rays can be described by geodesic deviation vectors (connecting vectors of
the observer's lightcone) from a central null ray connecting the center of
the source's shape to the observer.  }

\label{fig:distIIb}  
\end{figure} 

\newpage
\begin{figure}
\centerline{\psfig{figure=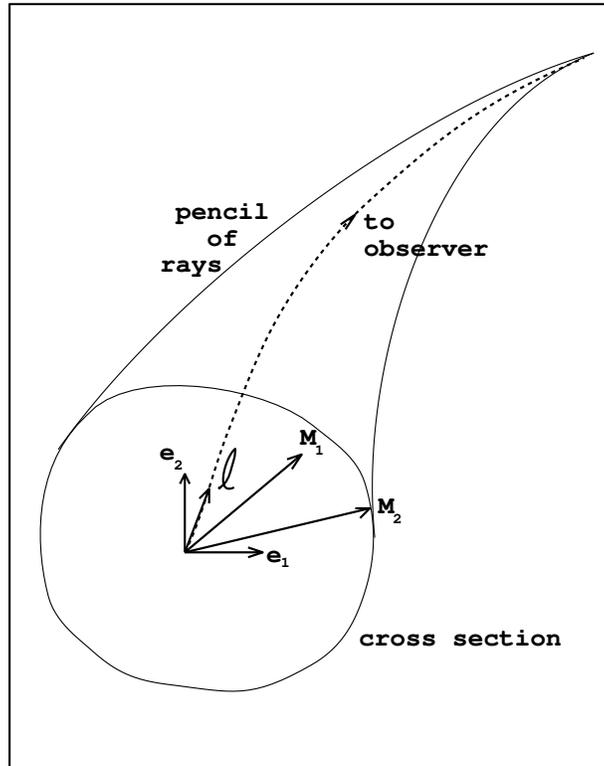,height=4in,angle=0}}

\caption{Vectors associated with the null geodesic that joins the source to
the observer.  The vector $\ell$ is tangent to the null geodesic.  The
vectors $(e_1,e_2)$ are spacelike, orthogonal to $\ell$, orthonormal and
parallel progagated along the geodesics. The vectors $(M_1,M_2)$ are
linearly independent Jacobi fields which coincide with $(e_1,e_2)$ at the
location of the source (not shown). }

\label{fig:distIIc}  
\end{figure} 

\end{document}